\documentclass[preprint]{imsart}

\RequirePackage[OT1]{fontenc}
\RequirePackage{amsthm,amsmath}
\RequirePackage{natbib}
\RequirePackage[colorlinks,citecolor=blue,urlcolor=blue]{hyperref}
\usepackage{graphicx}
\usepackage{MnSymbol}

\startlocaldefs
\numberwithin{equation}{section}
\theoremstyle{plain}

\endlocaldefs

\begin{document}

\begin{frontmatter}
\title{Bayesian Model Averaging for the X-Chromosome Inactivation Dilemma in Genetic Association Study}
\runtitle{BMA for the XCI Dilemma in GWAS}
\thankstext{T1}{To whom correspondence should be addressed.}

\begin{aug}
\author{\fnms{Bo} \snm{Chen}},
\author{\fnms{Radu} \snm{V. Craiu}\thanksref{T1}\ead[label=e2]{craiu@utstat.toronto.edu}}
\and
\author{\fnms{Lei} \snm{Sun}\thanksref{T1}
\ead[label=e3]{sun@utstat.toronto.edu}}

\runauthor{B. Chen et al.}

\affiliation{Department of Statistical Sciences, University of Toronto}

\address{Department of Statistical Sciences\\
University of Toronto\\
Toronto, ON M5S3G3, Canada\\
\printead{e2}\\
\printead{e3}\\
}

\end{aug}

\begin{abstract}
{X-chromosome is often excluded from the so called `whole-genome' association studies due to its intrinsic difference between males and females.  One particular analytical challenge is the unknown status of X-inactivation, where one of the two X-chromosome variants in females may be randomly selected to be silenced.  In the absence of biological evidence in favour of one specific model, we consider a Bayesian model averaging framework that offers a principled way to account for the inherent model uncertainty, providing model averaging-based posterior density intervals and Bayes factors.  We examine the inferential properties of the proposed methods via extensive simulation studies, and we apply the methods to a genetic association study of an intestinal disease occurring in about twenty percent of Cystic Fibrosis patients. Compared with the results previously reported assuming the presence of inactivation, we show that the proposed Bayesian methods provide more feature-rich quantities that are useful in practice.}

\end{abstract}


\end{frontmatter}

\section{Introduction}
\label{introduction}

In the search for genetic markers that are responsible for heritable complex human traits, whole-genome scans including genome-wide association studies (GWAS) and the next generation sequencing (NGS) studies have made tremendous progress; see {\texttt{www.genome.gov/gwastudies}} for the most recent summary of GWAS findings by the National Human Genome Research Institute \citep{welter14}.  The `whole-genome' nature of these studies, however, is often compromised by the omission of the X-chromosome \citep{heid10, teslovich10}.  In fact, it was found that ``{\it only 33\% (242 out of 743 papers)  reported including the X-chromosome in analyses}" based on the NHGRI GWAS Catalog \citep{wise13}.  The exclusion of X-chromosome from GWAS and NGS is due to it being fundamentally different between females and males.  In contrast to the 22 autosomal chromosomes where both females and males have two copies, females have two copies of X-chromosome (XX) while males have only one X coupled with one Y-chromosome (XY). Thus, statistical association methods well developed for analyzing autosomes require additional considerations for valid and powerful application to X-chromosome.  

Focusing on the single nucleotide polymorphisms (SNPs) as the genetic markers of interest here and without loss of generality, let $d$ and $D$ be the two alleles of a SNP and $D$ be the risk allele. An X-chromosome SNP in females has three possible (unordered) genotypes, $dd$, $dD$ and $DD$, in contrast to $d$ and $D$ in males.  Suppose each copy of the $D$ allele has an effect size of $\beta$ on the outcome of interest; this $\beta$ is the coefficient in linear regression for studying (approximately) normally distributed outcomes, or the log odds ratio in logistic regression for analyzing binary traits.  To ensure ``{\it dosage compensation for X-linked gene products between XX females and XY males}", X-chromosome inactivation (XCI) may occur so that one of the two alleles in females is randomly selected to be silenced \citep{gendrel11}.  In other words, the effects of  $dd$, $dD$ and $DD$ in females are now respectively 0, $\beta/2$ and $\beta$ on average after XCI vs.\ 0, $\beta$ and $2\beta$ without XCI.   However, without collecting additional biological data the status of XCI is unknown.

Previous work on developing association methods for X-chromosome SNPs mostly focused on issues other than XCI, including the assumptions of Hardy-Weinberg equilibrium (HWE) and equal allele frequencies or sample sizes between female and males \citep{zheng07, clayton08}.  In his classic review paper,  \cite{clayton09} also discussed analytical strategies for multi-population or family-based studies.  In each of these cases, either the XCI or no-XCI model is assumed, and naturally these methods work well only if the underlying assumption about the XCI status is correct  \citep{loley11, hickey11, konig14}.  

More recently, \citet{wang14} recognized the problem and proposed a maximum likelihood approach.  In essence, the proposed method calculates multiple association statistics for testing the effect of a X-chromosome SNP under XCI and no XCI models, then uses the maximum. To adjust for the inherent selection bias, the method uses a  permutation-based procedure to obtain the empirical distribution for the maximal test statistic and assess its significance.   
Although \citeauthor{wang14}'s method appears to be adequate in terms of association testing, in the presence of model uncertainty it is not clear how to construct a point estimate or confidence interval for effect size $\beta$, or,  what is a suitable measure of evidence for supporting one model over the other.   Thus, an alternative paradigm that directly accounts for the inherent model uncertainty is desirable.   

To close this gap, we propose a Bayesian approach that can handle in a principled manner the uncertainty about  the XCI status.   The use of Bayesian methods for genetic association studies is not new. \citet{stephens09} and  \cite{craiu14} provide reviews in the context of studying autosome SNPs.  Here we consider the posterior distributions generated from Bayesian regression models for analyzing X-chromosome SNPs under the XCI and no XCI assumptions.  We combine the estimates from the two models following the Bayesian model averaging (BMA) principle that  has long been recognized as a proper  method for incorporating model uncertainty in a Bayesian analysis \citep{draper,hoeting}.  We calculate the BMA-based highest posterior density (HPD) region for the parameter of interest.  The BMA posterior distribution is directly interpretable as a weighted average for $\beta$, averaged over the XCI and no XCI models with more weight given to the one with stronger support from the data. To rank multiple SNPs, we calculate Bayes factors comparing the averaged model with the null model of no association for each SNP.

In Section \ref{methods}, we present the theory of Bayesian model averaging for handling the X-chromosome inactivation uncertainty issue.  We first consider linear regression models for studying continuous traits where closed-form solutions can be derived.  We then discuss extension to logistic models for analyzing binary outcomes where Markov chain Monte Carlo (MCMC) methods are used for inference.  In this setting, the calculation of Bayes factors is no longer possible analytically so we implement numerical approximations that have been reliably used in computing ratios of normalizing constants. To facilitate methods comparison, we also provide an analytical solution for assessing significance of the maximum statistic in the spirit of \cite{wang14}, supplanting their permutation-based approach.  In Section \ref{simulation},  we conduct extensive simulation studies to evaluate the performance of the proposed Bayesian approach.  In Section \ref{realD}, we apply the method to a X-chromosome association study of meconium ileus, an intestinal disease present in Cystic Fibrosis patients, providing further evidence of method performance.  In Section \ref{discussion}, we discuss possible extensions and future work. 

\section{Methods}
\label{methods}

\subsection{Normally distributed outcomes}
\label{normal}

The methodology development here focuses on linear models, studying association relationship between a (approximately) normally distributed trait/outcome $Y$ and a X-chromosome SNP.  
Let $(dd,dD,DD)$ and $(d,D)$ be the genotypes of a SNP, respectively, for females and males.  For autosome or X-chromosome SNPs in females, genotypes $dd,dD$ and $DD$  are typically coded additively as 0, 1 and 2, representing the number of copies of a reference allele, assumed to be $D$ here.  Under the X-chromosome inactivation (XCI) assumption, one of the two alleles of a female is randomly selected to have no effect on the outcome.   Thus, the XCI and no XCI assumptions lead to two different coding schemes, respectively, $G_1$ and $G_2$ as summarized in Table \ref{coding}.

Let $Y$ be the vector of outcome measures of sample size $n$, and $G_k$ be the vector of genotype values for the $n$ individuals coded under model $M_k$, $k=1$ and $2$ as shown in Table \ref{coding}.  For each model $M_k$, we consider a linear regression model $Y=X_k\boldsymbol\theta_k+\epsilon_k$, where $X_k=(\mathbf{1}_{n},G_k)$ is the design matrix, 
 $\boldsymbol\theta_k=(\alpha_k, \beta_k)'$ and $\epsilon_k \sim N(0,\sigma^2I_n)$. Here $\beta_k$ represents the genetic effect of one copy of $D$ under model $M_k, k=1, 2$, accounting for the effects of other covariates $Z$s such as gender, age, smoking status and population information. For notation simplicity and without loss of generality for implementing the following Bayesian model average framework, $Z$s are omitted from the regression model.  The coding of 0.5 for genotype $dD$ under $M_1$ reflects the fact that the effect of $dD$ under the XCI assumption is the average of zero effect of $d$ (if $D$ was silenced) and $\beta$ effect of $D$ (if $d$ was silenced).  In addition, $\epsilon_1$ and $\epsilon_2$ have the same variance $\sigma^2I_n$ because both models are based on same response variable $Y$. 

Before we present the Bayesian approach, we make several important remarks here.  First, the regression model above studies the genotype of a SNP, thus it does not require the assumption of HWE; only methods based on allele counts are sensitive to the equilibrium assumption \citep{sasieni97}.  Similarly, allele-frequency affects only they efficiency of genotype-based association methods but not the accuracy. In addition, although other types of genetic architecture are possible, e.g. $dD$ and $DD$ having the same effect as in a dominant model or $dd$ and $dD$ having the same effect as in a recessive model, the additive assumption has its theoretical justification and sufficiently approximates many other models \citep{hill08}.   

\subsection{A Bayesian model averaging approach}
\label{BMA}
In practice, it is unknown which of the two models ($M_{1}$ XCI and $M_{2}$ no XCI ) is true.  Instead of performing inference based on only one of the two models or choosing the maximum one, the Bayesian model averaging (BMA) framework naturally aggregates information from both $M_{1}$ and $M_{2}$.   Central to BMA is the Bayes factor ($BF$) defined as
$$BF_{12}=\frac{P(Y|M_1)}{P(Y|M_2)},$$
where $P(Y|M_{k})=\int f(Y|\boldsymbol\theta,\sigma^2, M_k)\pi(\boldsymbol\theta|\sigma^2, M_k)\pi(\sigma^2|M_k)d\boldsymbol\theta d\sigma^2$ is the marginal probability of the data under model $M_k$.  Here we used the outcome variable $Y$ to denote  all available data; meaning should be clear from the context. 
 
We consider conjugate priors for  $\pi(\sigma^2|M_k$) and $\pi(\boldsymbol\theta|\sigma^2, M_k)$ for each model, $\pi(\sigma^2|M_k)=\pi(\sigma^2)=IG(a_0,b_0)$  where $IG(a_0,b_0)$ is the inverse gamma distribution with density function
$$p(\sigma^2)=\frac{b_0^{a_0}}{\Gamma(a_0)}(\sigma^2)^{-a_0-1}\exp\left(-\frac{b_0}{\sigma^2}\right).$$
As noted before,  $Y$ is common between $M_1$ and $M_2$ so the prior distributions of $\sigma^2$ for the two models are the same.  For  $\pi(\boldsymbol\theta|\sigma^2, M_k)=\pi(\boldsymbol{\theta}_{k})$, 
$$\pi(\boldsymbol{\theta}_{k})= N(\boldsymbol{\mu_0},\sigma^2\Lambda_{0k}^{-1}),$$ 
where $\Lambda_{0k}$ is the precision matrix \citep{wright08}.
For hyperparameter $\Lambda_{0k}$, we adopt the g-prior \citep{zellner86} that takes the form of $\Lambda_{0k}=\frac{\lambda}{n}X_k'X_k$.  We note that here the female component of $G_1$ is half of that of $G_2$.   Thus, if we na\"{i}vely use $\pi(\boldsymbol{\theta}_{k})= N(\boldsymbol{\mu_0},\sigma^2 \lambda I_2)$, this scaling factor can affect the Bayes factor and the ensuing model average quantities; the model with smaller covariate values is always preferred even if rescaling is the only difference. We discuss further  in Section \ref{discussion} the importance of using the g-prior form in this setting.

When estimating the posterior distribution of $\boldsymbol\theta$ under each model, the effect of the precision parameter $\lambda$ is minimal,  but this is not true for inferring whether  $\beta=0$ or not using the Bayes factor.  For the latter purpose, following the recommendations in \cite{kass} we use $\lambda=1$.  For other hyperparameters, naturally ${\boldsymbol{\mu_0}}=\mathbf 0$ unless there is prior information about association between the SNP under the study and the trait of interest.  In the absence of additional information for $\sigma^2$, we let $a_0=b_0=0.1$; setting $a_0=b_0=0$ in simulation studies did not lead to noticeable numerical difference compared to $a_0=b_0=0.1$. 

The likelihood function is defined by  $f(Y|\boldsymbol\theta, \sigma^2, M_k) \sim N(X_k \boldsymbol{\theta},\sigma^2I_n)$, which yields a normal-inverse-gamma posterior distribution, and the corresponding marginal distributions of $\boldsymbol{\theta}$ and $\sigma^2$ can be derived.  Specifically, $\pi(\boldsymbol{\theta},\sigma^2|Y, M_k)$, the posterior distributions for  $(\boldsymbol{\theta},\sigma^2)$ under each model $M_k$, is a multivariate t distribution with $2a$ degrees of freedom (df henceforth), location parameter $\boldsymbol{\mu_k}$ and scale parameter $\frac{b_k}{a}\Lambda_k^{-1}$, i.e., density function
$$\pi(\boldsymbol{\theta}|Y,M_k) \propto [1+\frac{(\boldsymbol{\theta}-\boldsymbol\mu_k)'\Lambda_k(\boldsymbol{\theta}-\boldsymbol\mu_k)}{2b_k}]^{-\frac{2a+2}{2}},$$
and the posterior of $\sigma^2$ is
$\pi(\sigma^2|Y, M_k) =IG(a,b_k),$
where
$$\Lambda_k=X_k'X_k+\Lambda_{0k} \: \:\: (\Lambda_{0k}=\frac{\lambda}{n}X_k'X_k),$$
$$\boldsymbol\mu_k=\Lambda_k^{-1} (\Lambda_{0k}\boldsymbol{\mu_0}+X_k'Y),$$
$$a=a_0+\frac{n}{2}, \mbox{ and } b_k=b_0+\frac{1}{2}(Y'Y+\boldsymbol{\mu_0}'\Lambda_{0k}\boldsymbol{\mu_0}-\boldsymbol\mu_k'\Lambda_k\boldsymbol\mu_k).$$

Focusing on the primary parameter of interest here, we extract the slope coefficient $\beta$ from the posterior of $\boldsymbol{\theta}=(\alpha, \beta)$ under each model $M_k$.  If we let $\mu_{k2}$ be the second element of $\boldsymbol\mu_k$, and $(\Lambda_k^{-1})_{22}$ be the $(2,2)_{th}$ entry in $\Lambda_k^{-1}$, we obtain that $\beta$ has univariate t distribution with $2a$ df and  $\mu_{k2}$ and $\frac{b_k}{a}(\Lambda_k^{-1})_{22}$, respectively, as the location and scale parameters, i.e. 
\begin{equation}
\label{beta.posterior}
\pi(\beta|Y,M_k)=\mu_{k2}+t_{2a}\sqrt{\frac{b_k}{a}(\Lambda_k^{-1})_{22}},
\end{equation}
 where $t_{2a}$ is the standard t distribution with  $2a$ df.
The normalizing constant  for the posterior under model $M_{k}$ is then
$$P(Y|M_k)=\frac{f(Y|\boldsymbol\theta,\sigma^2, M_k)\pi(\boldsymbol\theta|\sigma^2, M_k)\pi(\sigma^2|M_k)}{\pi(\boldsymbol{\theta},\sigma^2|Y, M_k)}
=\frac{1}{(2\pi)^{n/2}}\sqrt{\frac{|\Lambda_{0k}|}{|\Lambda_k|}}\frac{b_0^{a_0}\Gamma(a)}{b_k^a\Gamma(a_0)},$$
which leads to the Bayes factor between $M_{1}$ and $M_{2}$ as
\begin{equation}
BF_{12}=\sqrt{\frac{|\Lambda_2|}{|\Lambda_1|} \times \frac{|\Lambda_{01}|}{|\Lambda_{02}|}}\left (\frac{b_2}{b_1}\right)^a.
\label{bf12}
\end{equation}

The BMA of two models takes the form of \citep{hoeting} 
\begin{eqnarray*}
\pi(\boldsymbol{\theta},\sigma^2|Y)&=&P(M_1|Y)\pi(\boldsymbol{\theta},\sigma^2|Y, M_1)+P(M_2|Y)\pi(\boldsymbol{\theta},\sigma^2|Y, M_2).
\end{eqnarray*}
Let $P(Y)$ be the marginal probability of the data obtained after averaging over both models,
\begin{equation}
P(Y)=P(Y|M_{1})P(M_{1})+P(Y|M_{2})P(M_{2}).
\label{marg}
\end{equation} 
In the absence of prior information, it is customary to assume equal prior probabilities for the two models, i.e. $P(M_1)=P(M_2)=0.5$.  Therefore we have 
$$\pi(\boldsymbol{\theta},\sigma^2|Y)= \frac{P(Y|M_1)P(M_1)}{P(Y|M_1)P(M_1)+P(Y|M_2)P(M_2)} \pi(\boldsymbol{\theta},\sigma^2|Y, M_1)$$
$$+\frac{P(Y|M_2)P(M_2)}{P(Y|M_1)P(M_1)+P(Y|M_2)P(M_2)} \pi(\boldsymbol{\theta},\sigma^2|Y, M_2)$$
\begin{equation}
=\frac{BF_{12}}{1+BF_{12}} \pi(\boldsymbol{\theta},\sigma^2|Y, M_1)+ \frac{1}{1+BF_{12}} \pi(\boldsymbol{\theta},\sigma^2|Y, M_2).
\label{bmaposterior}
\end{equation}
Note that the posterior distribution $\pi(\boldsymbol{\theta},\sigma^2|Y)$, which we call {\it BMA posterior}, is a mixture of the two posterior distributions resulting from models $M_1$ and $M_2$. 
Because it is not obtained  from a given sampling distribution and a particular prior, it may not be a canonical posterior. 

The BMA posterior relies on the Bayes factor as the weighting factor, favouring one model over using weights based on $BF_{12}$.  Given an established association, we expect the Bayes factor provide evidence supporting one of the two models. Intuitively, if $BF_{12}>1$ then we have more support for $M_1$ from the data and vice versa when $BF_{12}<1$.  For the priors considered here,  we show in the Supplementary Materials that when data was generated from $M_1$, $Y=X_1\boldsymbol{\theta}_1+\epsilon_1$, ${BF_{12}}  \overset{p}{\to} \infty$ as $n \to \infty$ for any values of the hyperparameters, and similarly when $Y=X_2\boldsymbol{\theta}_2+\epsilon_2$, $BF_{12} \overset{p}{\to} 0$.  This is also consistent with our empirical observations from simulation studies, supporting the use of Bayes factor for model selection in this setting.

\subsection{BMA-based highest posterior density interval for the genetic effect of a SNP}
\label{hpd}

There are multiple ways to assess the genetic effect of a SNP based on the posterior distribution of $\beta$. The  simpler approach is to use the posterior mode of $\beta$ as a point estimate.  The highest posterior density (HPD) region however provides more information with an interval estimate.  To calculate BMA-based HPD, we note that the posterior density of $\beta$ from each of the $M_1$ and $M_2$ models is a univariate t with location and scale parameters as specified in equation (\ref{beta.posterior}).  The BMA posterior of $\beta$ is therefore a mixture of two known t distributions with the mixture proportion depending on $BF_{12}$. It is thus possible to calculate the exact HPD region for $\beta$.  

A $(1-\alpha)\%$ HPD is defined as $R(c_{\alpha})=\{\beta: \pi(\beta|Y) \geq c_{\alpha}\}$, where $\pi(\beta|Y)$ is the BMA posterior density of $\beta$ and $c_\alpha$ is the threshold such that the area under the posterior density is $1-\alpha$. Depending on the similarity between the two posterior distributions corresponding to $M_1$ and $M_2$ for a given credible level $\alpha$, a BMA HPD region can be either one single interval or made up of two disconnected intervals.  In all examples we have studied the HPD region is  a single interval at $\alpha=0.05$. Specifically, let $\beta_l$ and $\beta_u$ to be the two solutions of $\pi^{-1}(c_{\alpha})$. The  $1-\alpha$  HPD region is then ($\beta_l, \beta_u$), where
$$\int_{\beta_l}^{\beta_u}\pi(\beta|Y)d\beta=1-\alpha,$$
\begin{equation}
\pi(\beta_l|Y)=\pi(\beta_u|Y)=c_{\alpha}.
\label{HPD}
\end{equation}
The closed form of $\pi(\beta|Y)$ is in fact available, thus we can solve the equations defined in (\ref{HPD}) numerically to find $c_{\alpha}$ as well as $\beta_l$ and $\beta_u$, using function \texttt{multiroot} in R package \texttt{rootSolve}.
Note that for notation simplicity, we use $\alpha$ here to denote the desired credible level; its  distinction from the intercept parameter, also denoted by $\alpha$, should be clear from the context.

\subsection{Assessing genetic effect and ranking multiple SNPs by Bayes factor}
\label{testing}

In Bayesian framework, the significance of  a SNP can be evaluated using Bayes factor \citep{kass,stephens09}.  In the presence of model uncertainty, we propose using the Bayes factor calculated by comparing the averaging model between $M_1$  and $M_2$ with the null model of no effect, $M_N$. 
Under the null model of $\beta=0$, let $X_N=\mathbf{1_n}$ be the corresponding design matrix.  Using the same prior distributions and hyperparameter values for the remaining parameters, $\sigma^2$ and $\alpha$,   the calculation of $P(Y|M_N)$ is then similar to that of $P(Y|M_1)$ and $P(Y|M_2)$ as described in Section \ref{BMA}.  Let 
$$BF_{1N}=\frac{P(Y|M_1)}{P(Y|M_N)}, \:\:\: BF_{2N}=\frac{P(Y|M_2)}{P(Y|M_N)}$$ 
be the Bayes factors comparing, respectively, the XCI $M_1$ and no XCI $M_2$ with the null model $M_N$, the Bayes factor for comparing the averaging model with the null model is defined as
$$BF_{AN}=\frac{P(Y|M_{1})P(M_{1})+P(Y|M_{2})P(M_{2})}{P(Y|M_N)}.$$
Because $P(M_1)=P(M_2)=0.5$ in our setting, we thus have 
\begin{equation}
BF_{AN}=\frac{1}{2}(BF_{1N}+BF_{2N}).
\label{bma-bf}
\end{equation}
The Bayes factor $BF_{AN}$ has similar asymptotic properties as $BF_{12}$. We show in the Supplementary Materials that in our setting if $\lambda>0$ (the precision parameter for $\beta$), then $BF_{AN}$ converges in probability to either 0 or $\infty$, depending on whether $\beta=0$ or not.   

In practice, besides assessing association evidence for a single SNP, scientists are often interested in ranking multiple SNPs from a whole-genome scan and selecting the top ones for follow up studies.  The (conservative) lower bounds of the HPD intervals (and the BFs) can be used for this purpose, and we demonstrate this in Section \ref{realD} where we rank over 14,000 X-chromosome SNPs studying their association evidence with meconium ileus in Cystic Fibrosis patients.  

\subsection{Binary outcomes}
When we measure binary responses, $M_1$ and $M_2$ are logistic regression models.  Assuming the prior $\boldsymbol{\theta}_k \sim N(\boldsymbol{\mu_0}, \Lambda_{0k}^{-1})$, the BMA framework described above can still be used although computational complexities arise due to the lack of conjugacy.  Given its superior performance \citep{choi13}, we  use the Polya-Gamma sampler of \citet{polson13} and the R package \texttt{BayesLogit} to draw samples from the posterior distributions under $M_{1}$ and $M_{2}$.  To obtain samples from the averaged model, we draw samples from $M_1$ with probability ${BF_{12}}/{(1+BF_{12})}$ and from $M_2$ with probability ${1}/{(1+BF_{12})}$ based on equation (\ref{bmaposterior}). And we use these samples to construct the $1-\alpha$ HPD interval via  the function \texttt{HPDinterval} in the R package \texttt{coda}. 

The calculation of $BF_{12}$ is based on the Bridge sampling method proposed by \citet{meng96} and further refined by \citet{gelman98} which we delineate below. 
Suppose we have $J$ posterior samples, $\boldsymbol{\theta_{kj}}$, from the two models, $k=1$ and $2$ and $j=1,...,J$.  For each parameter sample $\boldsymbol{\theta_{kj}}$, we can calculate the corresponding unnormalized posterior density based on the logistic model under the $M_1$ XCI assumption,
\begin{eqnarray*}
q_1(\boldsymbol{\theta_{kj}})&=&\pi(\boldsymbol{\theta_{kj}}|M_1)f(Y|\boldsymbol{\theta_{kj}},M_1) \\ &=&\pi_1(\boldsymbol{\theta_{kj}})\prod_{i=1}^{n}p_{1i}(\boldsymbol{\theta_{kj}})^{Y_i}(1-p_{1i}(\boldsymbol{\theta_{kj}}))^{1-Y_i},
\end{eqnarray*}
where $p_{1i}(\boldsymbol{\theta_{kj}})=[1+ \exp(-X_{1i}\boldsymbol{\theta_{kj}})]^{-1}$, 
and $X_{1i}$ is the $i_{th}$ row of  the design matrix $X_1$ that contains the genotype data coded under model $M_1$ for the $i_{th}$ individual. $\pi_1$ is the density function of $N(\boldsymbol{\mu_0}, \Lambda_{01}^{-1})$. Similarly we obtain
\begin{eqnarray*} 
q_2(\boldsymbol{\theta_{kj}})&=&\pi(\boldsymbol{\theta_{kj}}|M_2)f(Y|\boldsymbol{\theta_{kj}},M_2) \\ &=&\pi_2(\boldsymbol{\theta_{kj}})\prod_{i=1}^{n}p_{2i}(\boldsymbol{\theta_{kj}})^{Y_i}(1-p_{2i}(\boldsymbol{\theta_{kj}}))^{1-Y_i},
\end{eqnarray*}
where $p_{2i}(\boldsymbol{\theta_{kj}})=[1+\exp(-X_{2i}\boldsymbol{\theta_{kj}})]^{-1}$ under model $M_2$, and $\pi_2$ is the density function of $N(\boldsymbol{\mu_0}, \Lambda_{02}^{-1})$.  We then define the ratio of unnormalized densities as $l_{kj}=q_1(\boldsymbol{\theta_{kj}})/q_2(\boldsymbol{\theta_{kj}})$ and compute the Bayes factor iteratively. Specifically, we set $BF_{12}^{(1)}=1$ and compute at the $(t+1)_{th}$ iteration until convergence,
\begin{equation}
BF_{12}^{(t+1)}=\frac{\sum_{j=1}^{J}\frac{l_{2j}}{l_{2j}+BF_{12}^{(t)}}}{\sum_{j=1}^{J}\frac{1}{l_{1j}+BF_{12}^{(t)}}}.
\label{bridge}
\end{equation}

When comparing the averaged model vs. null model, the above procedure cannot be directly implemented to calculate $BF_{1N}$ and $BF_{2N}$, since the null model has different dimension of parameter $\boldsymbol{\theta}$.  Instead of finding the ratio of normalizing constants by the numerical method above, we find $P(Y|M_1)$, $P(Y|M_2)$ and $P(Y|M_N)$ by calculating the ratio between them and known quantities. The latter will be the normalizing constants corresponding to Gaussian approximations of the posterior distributions of interest. More precisely, we use the following steps:
\begin{itemize}
\item To calculate $P(Y|M_1)$, we approximate the posterior under $M_1$ using a bivariate normal distribution with independent components. So we find the sample mean and sample variance of posterior sample $\boldsymbol{\theta_{1j}}=(\alpha_{1j},\beta_{1j})$, which are $(\bar{\alpha}_1$, $\bar{\beta}_1)$ and
$\left(\begin{array}{cc}
                   s^2_{\alpha_1} & 0  \\
                   0 & s^2_{\beta_1} \\
           \end{array}\right)$.
\item We simulate $\alpha'_{1j}$ and $\beta'_{1j}$ from the above bivariate approximation to the posterior whose normalizing constant is $c_1=2\pi s_{\alpha_1}s_{\beta_1}$ and set $\boldsymbol{\theta'_{1j}}=(\alpha'_{1j}, \beta'_{1j})$.
\item We use the  iterative approach in equation \eqref{bridge} to compute the ratio of normalizing constants between the posterior under $M_1$ and the corresponding approximation, $BF_1={P(Y|M_1)}/{c_1}$. Since $c_1$ is known, we can easily derive the normalizing constant $P(Y|M_1)$.  
\item To calculate $P(Y|M_N)$, we repeat the procedure used for $P(Y|M_1)$ but this time the dimension of the parameter is one instead of two.
\item The unnormalized posterior density for $M_N$ is
\begin{eqnarray*} 
q_N(\theta_{Nj})&=&\pi(\theta_{Nj}|M_N)f(Y|\theta_{Nj},M_N) \\
&=&\pi_N(\theta_{Nj})\prod_{i=1}^{n}p_{N}(\theta_{Nj})^{Y_i}(1-p_{N}(\theta_{Nj}))^{1-Y_i},
\end{eqnarray*}
where $p_{N}(\theta_{Nj})=[1+\exp(-\theta_{Nj})]^{-1}$, and $\pi_N$ is the prior density of $N(0, \lambda^{-1})$.
\item We then use equation \eqref{bridge}  to compute  $BF_N={P(Y|M_N)}/{c_N}$, and we obtain $BF_{1N}$ as
$$BF_{1N}=\frac{BF_1}{BF_N} \times \frac{c_1}{c_N}.$$ 
\item We repeat the above steps for $M_2$ to calculate $BF_{2N}$.
\item Finally,  we use equation \eqref{bma-bf} to calculate $BF_{AN}$ by averaging $BF_{1N}$ and $BF_{2N}$.
\end{itemize}

\subsection{Revisit the maximum likelihood approach of \cite{wang14}}
\label{wang}

Let $Z_1$ and $Z_2$ be the frequentist's test statistics  for testing $\beta_k=0$ derived from the two regression models, $Y=\alpha_k+\beta_k G_k+\epsilon_k, k=1$ and $2$, respectively under the XCI $M_1$ and no XCI $M_2$ assumptions; $G_k$ are the corresponding genotype codings as shown in Table \ref{coding}.   The maximum likelihood approach of \cite{wang14}, in essence, uses $Z_{max}=max(|Z_1|,|Z_2|)$ as the test statistic and calculates the p-value of $Z_{max}$ empirically via a permutation-based procedure.  Instead of obtaining p-values using a permutation-based method as discussed in \cite{wang14}, we note that the significance of $Z_{max}$ can be obtained more efficiently.  Under the null hypothesis of no association for either linear or logistic regression, $Z_1$ and $Z_2$ have an approximate bivariate normal distribution, $N(\textbf 0, \Sigma)$ and $\Sigma=\left( \begin{array}{cc} 1 &\rho_{Z_1, Z_2}  \\ \rho_{Z_1, Z_2}  & 1 \end{array}\right)$, where conditional on the observed genotypes $G_1$ and $G_2$, $\rho_{Z_1|G_1, Z_2|G_2}=r_{G_1, G_2}$, where $r_{G_1, G_2}$ is the sample correlation of $G_1$ and $G_2$.  This principle has been used in another setting where for an un-genotyped SNP, instead of imputing the missing genotype data, the association statistic is directly inferred based on the association statistic at a genotyped SNP and the correlation between the two SNPs estimated from a reference sample \citep{lee13, pasaniuc14}. Thus, given the two genotype codings of each SNP, without permutation we can find the threshold value $z_{1-\alpha}$ for the maximum statistic $Z_{max}$ at the nominal type I error rate of $\alpha$. In application study below (Section \ref{realD}), for each of the $14,000$ or so SNPs analyzed, we will obtain the corresponding p-value using this method.

\section{Simulation Study}
\label{simulation}

We conduct simulation studies to evaluate the performance of the proposed Bayesian model averaging methods for studying both normally distributed traits and binary outcomes.  Here we focus on the performance quantities relevant to Bayesian methods, including the BMA HPD internals of  Section \ref{hpd} and the BMA BF of Section \ref{testing}.  We leave the ranking comparison with the frequentist method of \cite{wang14} to the application study in Section \ref{realD}.

\subsection{Simulation settings}

In our simulations, we vary the sample size $n$, proportion of males and frequencies of allele $D$ for males and females ($p_{m}$ and $p_{f}$ respectively).  In each case we first generate data for $G$, where we simulate female genotypes using  a multinomial distribution with probabilities of $(1-p_f)^2$, $2p_f(1-p_f)$ and $p_f^2$, respectively, for $dd$, $dD$ and $DD$, and we simulate male genotypes using a binomial distribution with probabilities of $(1-p_m)$ and $p_m$, respectively, for $d$ and $D$.

We then generate outcome data for $Y$ based on the simulated $G$ coded under the XCI $M_1$ or no XCI $M_2$ assumption, and various parameter values of the regression models.  For linear models we fix $\alpha=0$; the intercept parameter has negligible effects on result interpretation (e.g.\ $\alpha=1$ lead to similar conclusion). Without loss of generality, we also fix $\sigma^2=1$.  Under the null model, $\beta=0$ and $Y$ does not depend on the XCI and no XCI assumptions, i.e.\ $Y \sim N(0,\sigma^2I_n)$.  Under alternatives and for each $M_k$, method performance depends on both genetic effect size $\beta$ and allele frequencies $p_m$ and $p_f$, via the quantity $EV$, the variation of $Y$ explained by genotype, where $EV={Var(E(Y|G))}/{Var(Y)}$.  Although allele frequencies affect method performance as we will see in the application study below, fixing $EV$ instead of $\beta$ has the benefit of not requiring specification of the relationship between $\beta$ and allele frequencies (e.g.\ variants with lower frequencies tend to have bigger effects or smaller effects, vs.\ the two parameters are independent of each other); \citet{andriy14} explored this in a frequentist setting for jointly analyzing multiple autosome SNPs.   For linear models, it is easy to show that  $EV={\beta^2\sigma^2_G}/{(\beta^2\sigma^2_G+\sigma^2)}$, where $\sigma^2_G$ is the variance of $G$ depending on $p_m$ and $p_f$. Thus, for a given $EV$ value we obtain 
$\beta=\sigma/\sigma_G \cdot \sqrt{EV/(1-EV)}$ 
for different choices of $p_m$ and $p_f$ and codings of $G$ for $M_k$, $k=1$ and 2. We then simulate $Y$ for continuous outcomes from $N(X_k\boldsymbol{\theta},\sigma^2I_n)$ based on $\boldsymbol{\theta}=(\alpha, \beta)$ and $X_k=(\mathbf{1}_{n},G_k)$.

For studying binary outcomes using logistic regression, we assume the typical study design of equal numbers of cases and controls. Under the null of $\beta=0$, we randomly assign $Y=0$ to half of the sample and $Y=1$ to the other half.  Under alternatives, the derivation of $\beta$ given $EV$ and allele frequencies is a bit more involved, and we outline the details in the Supplementary Materials.  We then simulate $Y$ from $Bin(n^*,(1+\exp(-X_k\boldsymbol{\theta}))^{-1})$, $n^*>n$, until $n/2$ numbers of cases and controls are generated.

To summarize, the parameters involved in the simulation studies include the sample size ($n$ and the proportion of males), allele frequencies in males and females  ($p_{m}$ and $p_{f}$), the variation of $Y$ explained by genotype ($EV$ and in turn $\beta$; without loss of generality (w.l.g.), $\alpha=0$, $\sigma^2=1$), as well as equal numbers of cases and controls for studying binary traits.  In the following, we show representative results when $n=1000$  (and assuming the proportion of males is half), $EV=0.005$ or $0.01$, and $p_m$ and $p_f$ ranging from 0.1 to 0.9 where the two frequencies do not have to be equal but have to be both $\geq 0.5$ or $\leq 0.5$; in practice it is unlikely that the difference in allele frequencies is so big that the reference alleles in males and females differ.  The number of MCMC samples for analyzing each binary dataset is $J=1000$. 

\subsection{Performance of the Bayesian methods}

We provide BMA-based HPD intervals and their corresponding $BF_{AN}$ on the $log_{10}$ scale for 50 independent data replicates simulated under different conditions for logistic regression models in Figures 1 and 2; results for linear models are provided in the Supplementary Materials.  The intervals are sorted by their lower bounds.  In the left-panel of each figure, the blue dashed line marks $\beta=0$ and the red solid line marks the true value of $\beta$; two lines overlap under the null.  In the right panel, the blue dashed line marks $log_{10}(BF_{AN})=0$, and the red solid line marks the conventional threshold of $log_{10}(BF_{AN})=1$ for declaring strong evidence favouring one model over the other \citep{kass}.

\begin{figure}[h]
\begin{center}\includegraphics[scale=0.25]{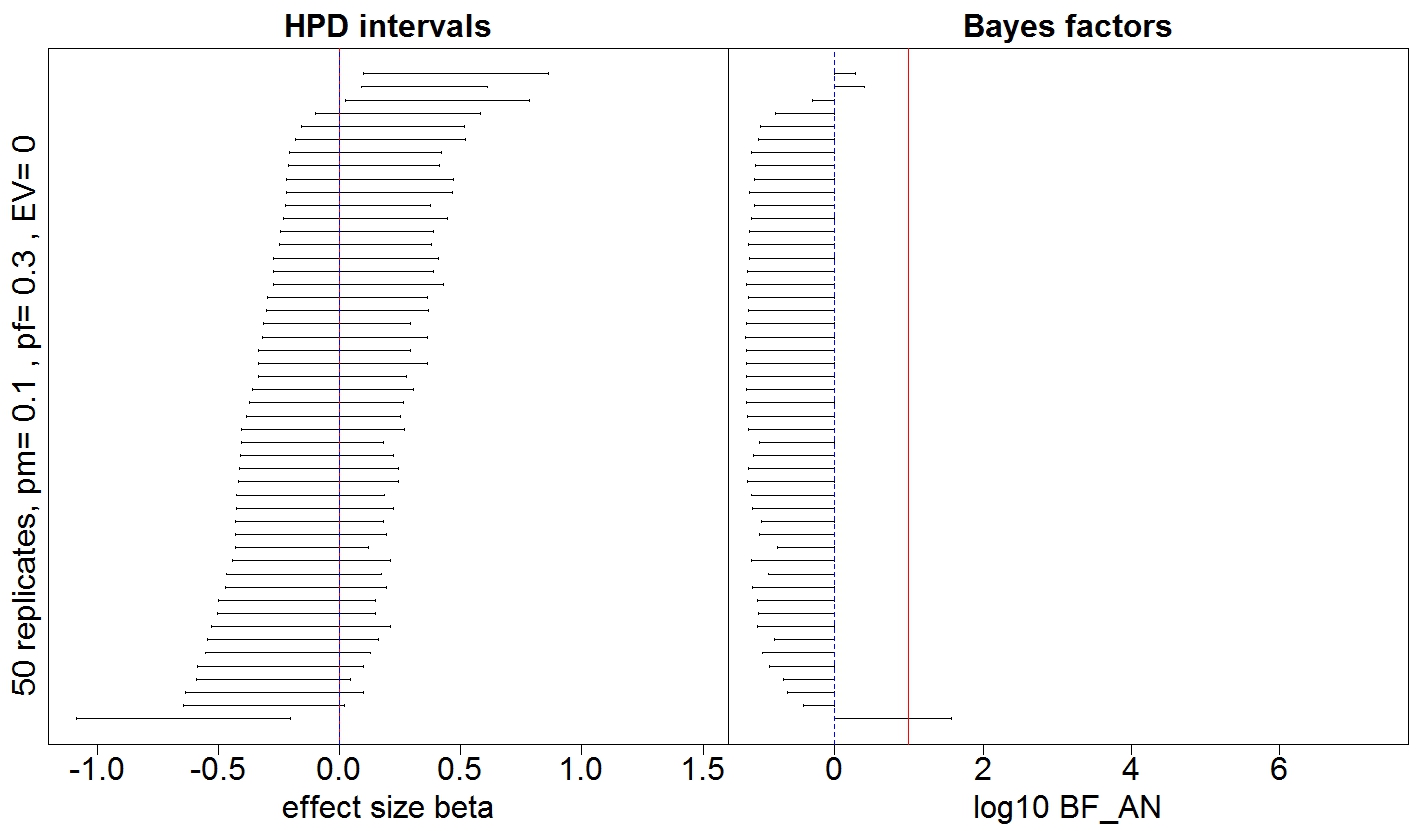}\end{center}
\begin{center}\includegraphics[scale=0.25]{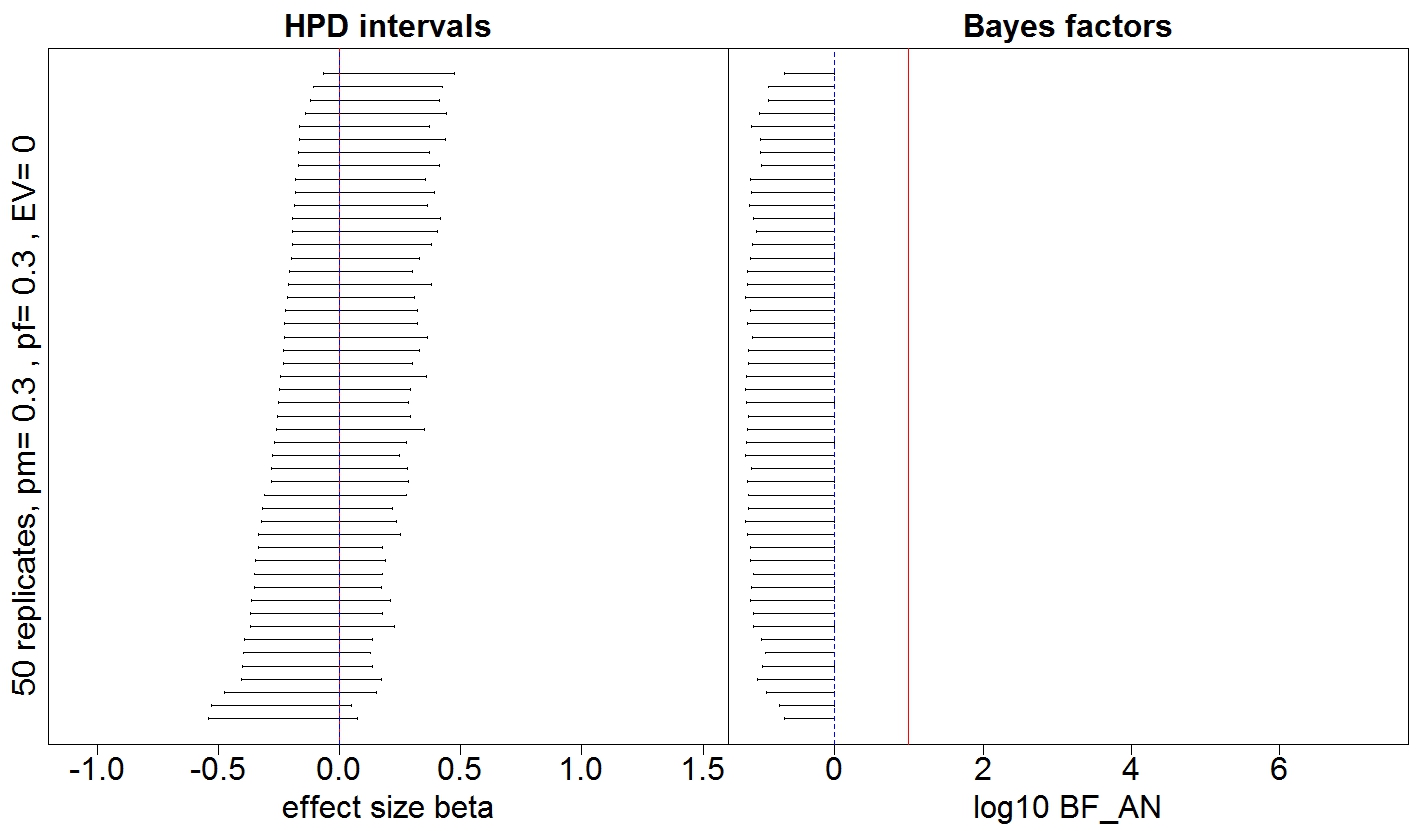}\end{center}
\caption{BMA-based HPD intervals for $\beta$ (left panel) and corresponding $log_{10}(BF_{AN})$ (right panel) for 50 replicates simulated based on null $M_N$ {\bf logistic model}.  Intervals are sorted by their lower bounds. 
Male and female allele frequencies are $(0.1, 0.3)$ (top panel) and $(0.3, 0.3)$ (bottom panel).  In the left panel, the blue dashed line and red solid line overlap at the null value of $\beta=0$.  In the right panel, the blue dashed line and red sold line mark, respectively, $log_{10}(BF_{AN})=0$ and $=1$.}  
\label{result logistic null}
\end{figure}

Figures \ref{result logistic null} present the results under the null of no association where $EV=0$ (i.e.\ $\beta=0$). The top panel is for unequal male and female allele frequencies at ($p_m, p_f$)=(0.1, 0.3), and the bottom panel is for ($p_m, p_f$)=(0.3, 0.3); results for other allele frequency values (e.g. (0.5, 0.3), (0.5, 0.7), (0.7, 0.7) and (0.9, 0.7)) are similar and provided in the Supplementary Materials. We note that although the HPD intervals do not have the same coverage interpretation as CI, just over 95\% of the HPD intervals contain zero.  Similarly, most of the values are less than zero, i.e. $BF_{AN}<1$; additional simulations show that $BF_{AN}$ decrease as the sample size increases under the null.

\begin{figure}[h]
\begin{center}\includegraphics[scale=0.25]{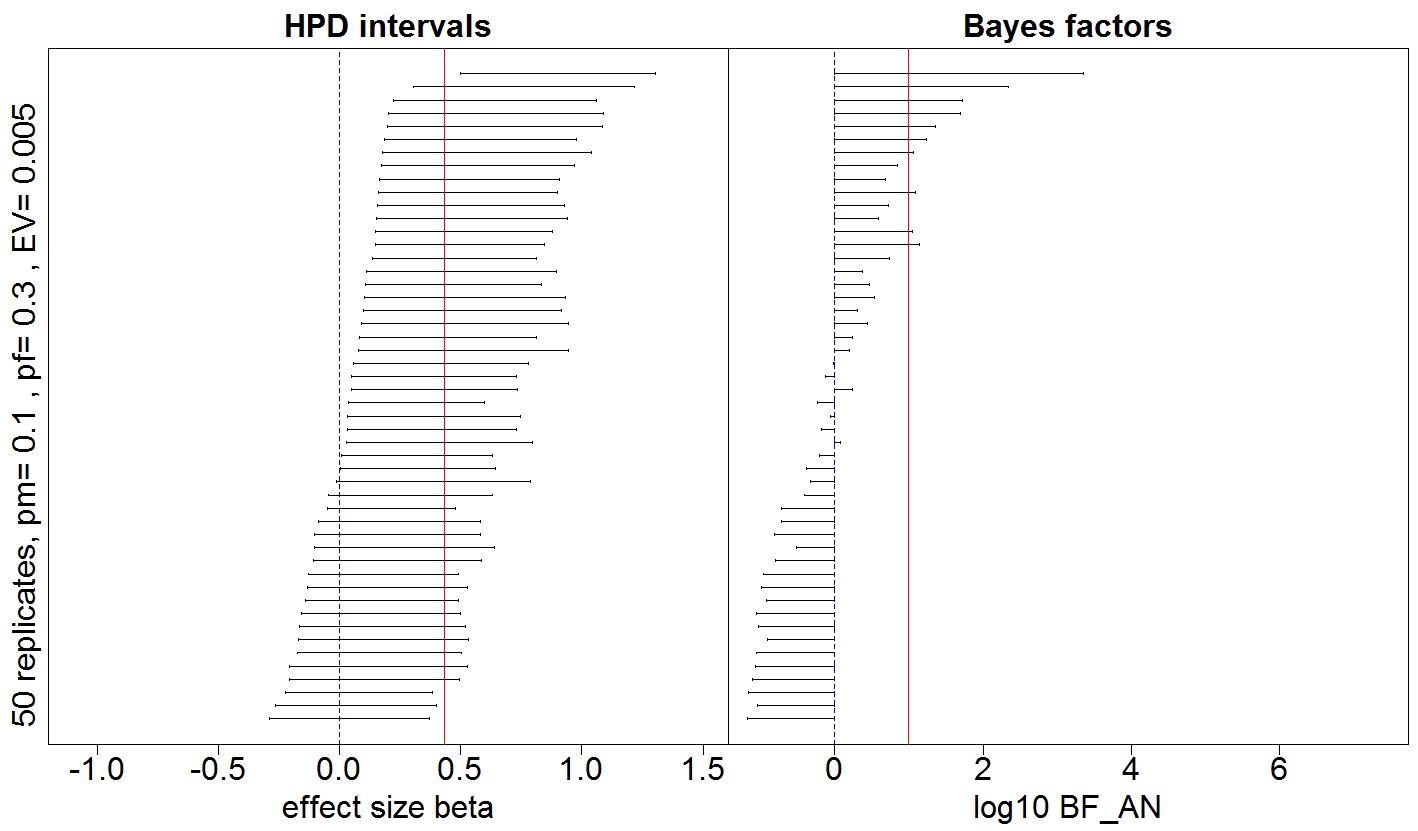}\end{center}
\begin{center}\includegraphics[scale=0.25]{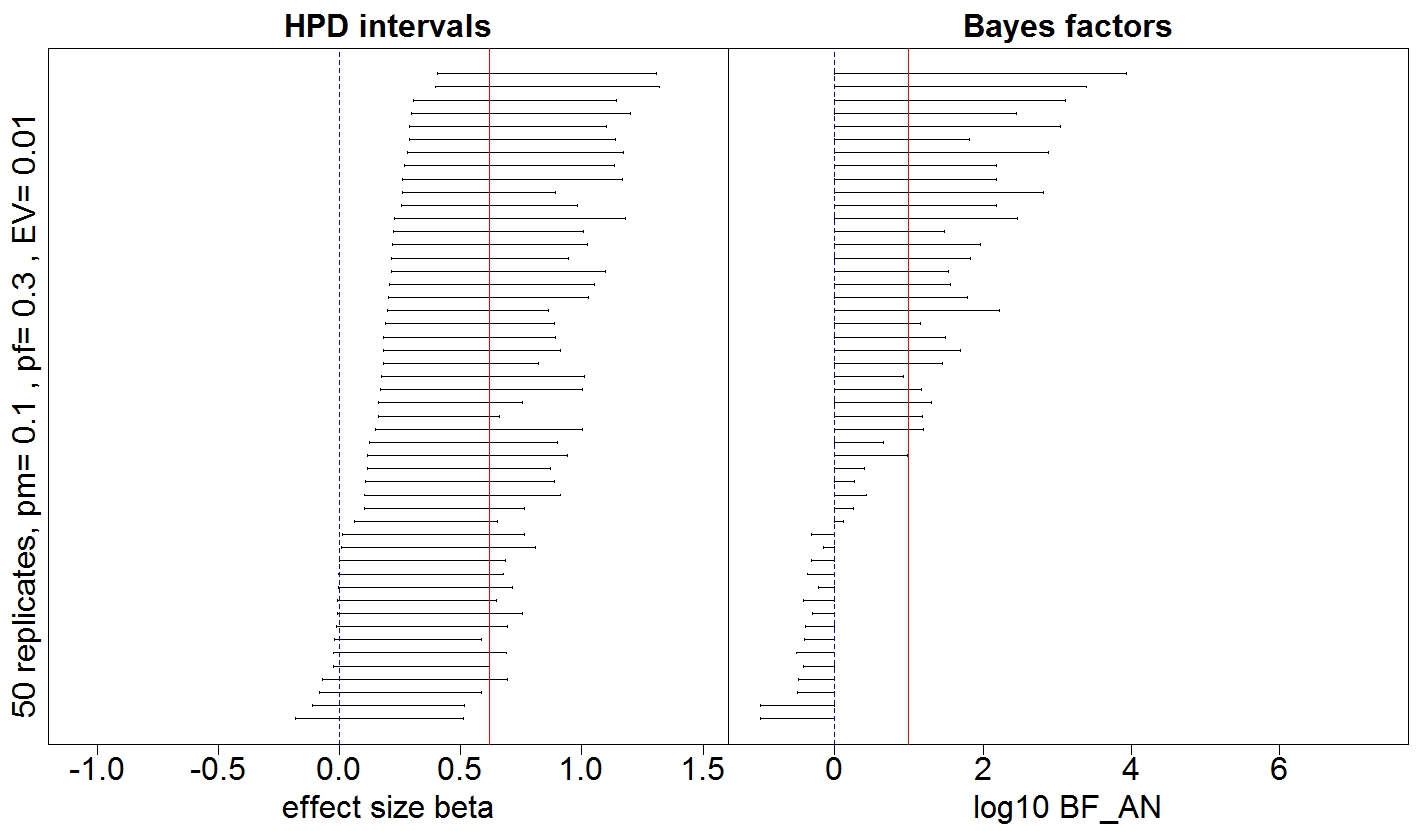}\end{center}
\caption{BMA-based HPD intervals for $\beta$ (left panel) and corresponding $log_{10}(BF_{AN})$ (right panel) for 50 replicates simulated based on XCI $M_1$ {\bf logistic model}.  Intervals are sorted by their lower bounds. 
Male and female allele frequencies are $(0.1, 0.3)$.  In the left panel, the blue dashed line marks the null value of $\beta=0$, and the red solid line marks the true value of $\beta$, determined so that $EV=0.005$ (top panel) and $EV=0.01$ (bottom panel).  In the right panel, the blue dashed line and red sold line mark, respectively, $log_{10}(BF_{AN})=0$ and $=1$.}  
\label{result logistic alter}
\end{figure}

Figures \ref{result logistic alter} presents the results under different alternatives, where ($p_m, p_f$)=(0.1, 0.3) and data are simulated from the XCI $M_1$ model, but $EV$ varies and $EV$=0.005 for the top panel and $EV$=0.01 for the bottom; results for other parameter values and data simulated from $M_2$ are similar and are included in the Supplementary Materials. It is clear that as $EV$ increases, the performance of the proposed Bayesian methods increases.

\section{Application Study}
\label{realD}

\cite{sun12} performed a whole-genome association scan on meconium ileus, a binary intestinal disease occurring in about 20\% of the individuals with Cystic Fibrosis. Their GWAS included X-chromosome but assumed the inactivation $M_1$ model.  They identified a gene called {\it SLCA14} to be associated with meconium ileus, and in their Table 2 they reported p-values in the range of $10^{-12}$, $10^{-8}$ and $10^{-6}$, respectively, for $rs3788766$, $rs5905283$ and $rs12839137$ from the region.  We revisit this data by applying  the maximum likelihood approach (or the minimal p-value of the XCI $M_1$ and no XCI $M_2$ models) and the proposed Bayesian model average method.

The data consists of $n=3199$ independent CF patients, and there are slightly more males ($n_m=1722$, 53.8\%) than females ($n_f=1477$).  Among the study subjects, 574 are cases with meconium ileus and 2625 are controls, and the rates  of meconium ileus ($17.7\%$ vs.\ $18.3\%$) do not appear to differ between the male and female groups. Genotypes are available for 14280 X-chromosome SNPs, but 60 are monomorphic (no variation in the genotypes). Thus association analysis is performed between each of the 14220 X-chromosome SNPs and the binary outcome of interest. That is, 14220 p-values and 14220 BMA BFs and HPD intervals are calculated and investigated.  By convention, for each SNP we assume the minor allele as the risk allele $D$ and we use the two coding schemes as described in Table \ref{coding} under the $M_1$ and $M_2$ models.  

\begin{figure}[h]
\begin{center}\includegraphics[scale=0.19]{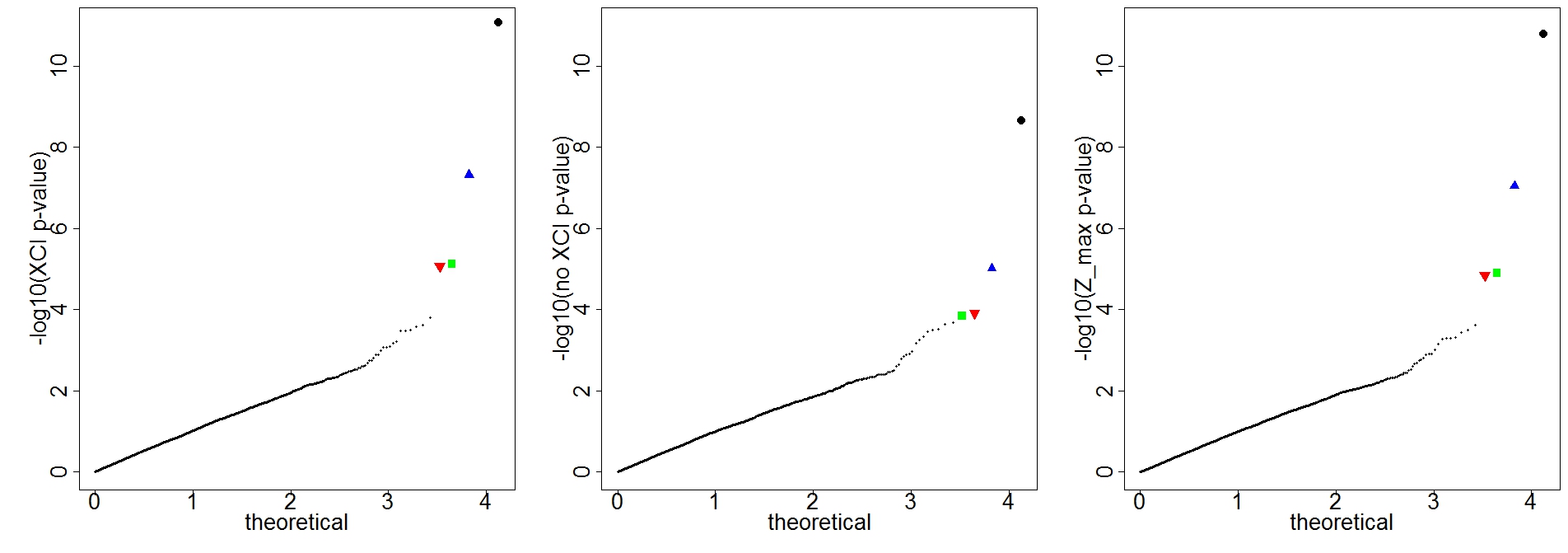}\end{center}
\caption{QQplots of $-log_{10}$ p-values of analyzing association evidence between 14220 X-chromosome SNPs and meconium ileus in 3199 Cystic Fibrosis patients, under the XCI $M_1$ assumption (left), the no-XCI $M_2$ assumption (middle) and using $Z_{max}$ (right).  Black circle $\bullet$ for $rs3788766$, blue up-pointing triangle {\color{blue}$\blacktriangle$} for $rs5905283$, green square {\color{green}$\blacksquare$} for $rs12839137$ and red down-pointing triangle {\color{red}$\blacktriangledown$} for $rs5905284$.}
\label{qqplots}  
\end{figure}

Figure \ref{qqplots} shows the QQplots of p-values obtained using the frequentist framework.  The left graph is under the $M_1$ inactivation assumption as in the original analysis of \citep{sun12}.  The middle one is under the $M_2$ no inactivation assumption.  And the right one is based on the minimal p-values adjusted for selecting the best of the two models using the asymptotic approximation as discussed in Section \ref{wang}.  As expected, most of the SNPs are from the null, but there are four clear outliers/signals with evidence for association with meconium ileus regardless of the methods used.  The overall consistency between the $M_1$ and $M_2$ models is the result of high correlation between $Z_1$ and $Z_2$; we discuss this point further in Section \ref{discussion} below.
Contrasting the left graph with the middle one in  Figure \ref{qqplots} shows that the XCI $M_1$ assumption lead to smaller p-values for these four SNPs than the no XCI $M_2$ assumption. Table \ref{real1} provides the corresponding minor allele frequencies (MAF, pooled estimates because sex-specific estimates are very similar to each other), log OR estimates and p-values for these four SNPs.

\begin{figure}[h]
\begin{center}\includegraphics[scale=0.22]{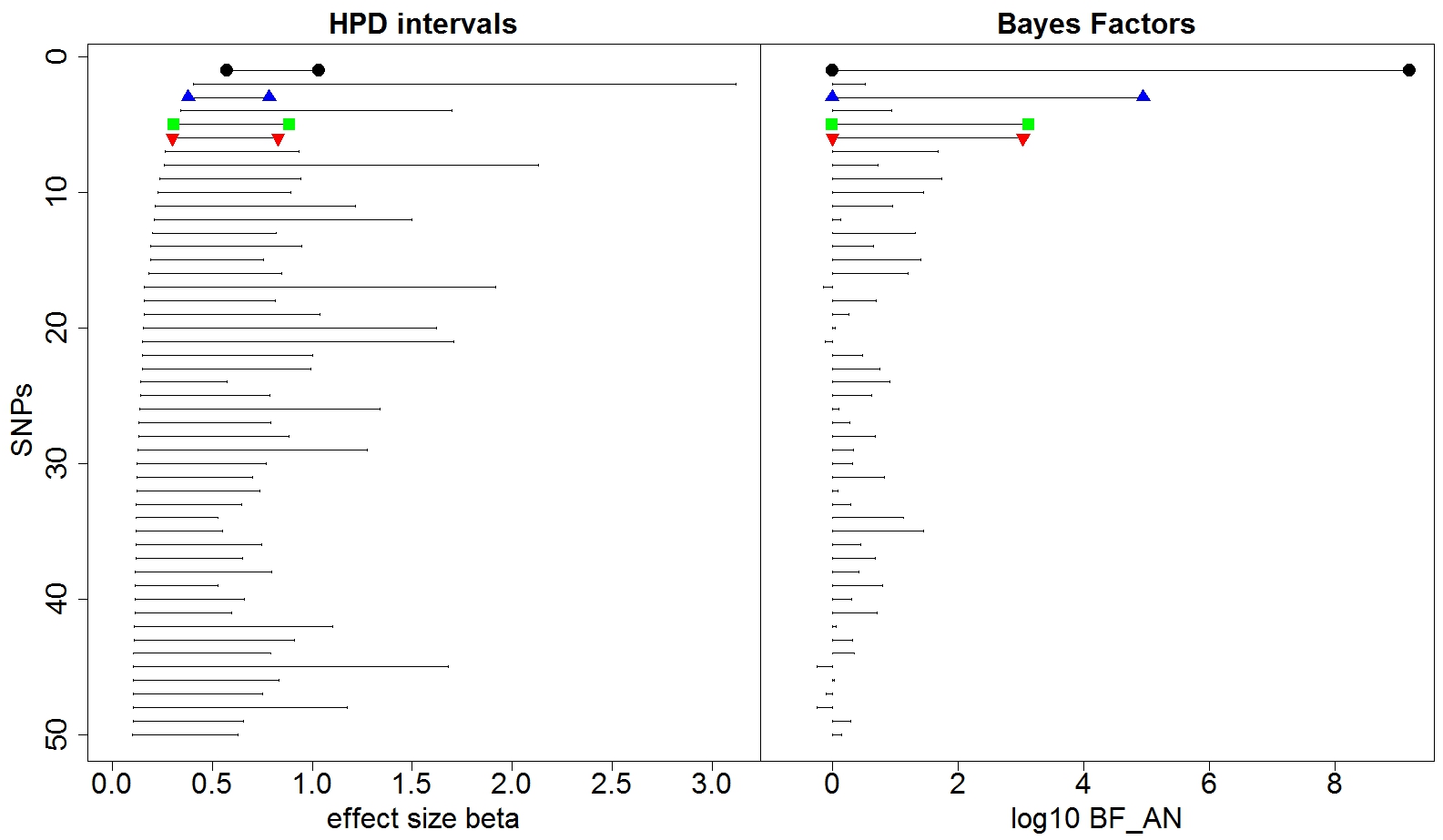}\end{center}
\caption{BMA-based HPD intervals for $\beta$ and corresponding $log(BF_{AN})$ for 50 top ranked SNPs, selected from analyzing association evidence between 14220 X-chromosome SNPs and meconium ileus in 3199 Cystic Fibrosis patients. SNP are ordered based on their lower bounds of the HPD intervals.  The four top SNPs identified by p-values are marked here using the same symbol: black circle $\bullet$ for $rs3788766$, blue up-pointing triangle {\color{blue}$\blacktriangle$} for $rs5905283$, green square {\color{green}$\blacksquare$} for $rs12839137$ and red down-pointing triangle {\color{red}$\blacktriangledown$} for $rs5905284$.}
\label{hpd real}  
\end{figure}

Figure \ref{hpd real} present the Bayesian results for top 50 ranked SNPs.  Similarly to the presentation of the simulation results in Section \ref{simulation}, in the left graph SNPs are ranked based on the lower bounds of their BMA-based HPD intervals, and the corresponding $log_{10}(BF_{AN})$ values are provided one the right. Table \ref{all SNPs} provides more detailed results for top 15 SNPs including the $BF_{12}$ comparing $M_1$ with $M_2$.  Note that for ease of presentation and without loss of generality we mirror all negative intervals to positive ones. 

Two important remarks can be made here.  First, the proposed Bayesian method clearly identifies the four SNPs suggested by the p-value approach. Second, the Bayesian framework in this setting provides more feature-rich quantities, and it pinpoints additional SNPs that merit follow-up studies.  Note that although p-values lead to similar rankings between the two models themselves, they could miss potentially important SNPs. 
Taking $rs12689325$ as an example,  this SNP will not be identified with a rank of 331 based on the p-value of $0.0268$, the adjusted minimum of p-values calculated under $M_1$ and $M_2$.   However, this SNP ranks second based on the (conservative) lower bounds of the BMA-based HPD intervals averaged over $M_1$ and $M_2$ (Figure \ref{hpd real} and Table \ref{all SNPs}).  The wide BMA HPD interval is a result of small MAF ($1.3\%$) coupled with a moderate effect size.  Given a trait of interest in practice, if genetic etiology implies the involvement of rare variants, the Bayesian results suggest that this SNP warrants additional investigation. 

\section{Discussion}
\label{discussion}

We propose a Bayesian  approach to address the ambiguity involved in GWAS and NGS studies of SNPs situated on the X-chromosome. Depending on whether X-inactivation takes place or not, there are two regression models that can be used to explore the genetic effect of a given SNP on the phenotype of interest.  The proposed method allows us to produce posterior-based inference that incorporates the uncertainty within and between genetic models. While the former is quantified by the posterior distribution under each model, the latter can be properly accounted for by considering a weighted  average of the model-specific estimators. Following the Bayesian paradigm, the weights are proportional to the Bayes factor comparing the two competing models. The asymptotic properties of the Bayes factors considered in this paper for linear models   are included  in the Supplementary Materials.  In the binary response case, the theoretical study is difficult due to the intractable posteriors, but the Monte Carlo estimators exhibit good properties in all the numerical studies performed.

The use of g-priors in this study setting is essential in that it allows us to avoid the effect of covariate rescaling on the Bayes factors, yet maintain results interpretation.  In regression models, we know that the effect size $\beta$ is inversely proportional to the size of the covariate value/genotype coding.  Given a set of data, using $X/2$ or $X$ should lead to identical inference.   However, without g-priors, a model with smaller covariate value would be preferred based on $BF$.  In our setting, the female component of the design matrix under the XCI $M_1$ coding is only half of that no XCI $M_2$ coding; male codings are the same for the two models.  Consider  the null case of $\beta=0$ when the two competing models are identical. Using $\Lambda_{0k}=\lambda I$ for the precision of $\boldsymbol{\theta_k}$, we see that 80\% of $BF_{12}$ are greater than one, suggesting $M_1$ is preferred simply because of its smaller genotype coding.  One statistical solution is to rescale the design matrix prior to the Bayesian inference. However, it is important to note that the coding difference for females is driven by a specific biological consideration, thus rescaling leads to difficulties in results interpretation.  Instead, we use a g-prior in Section \ref{methods}. Indeed, simulation results for the null case show that  $BF_{12}>1$  in about 50\% of replicates, indicating proper calibration.

The application has shown that the two models/assumptions lead to similar results, and one may argue that the practical benefits of using BMA-based inference is limited.  However, It's important to note that while the wrong model may still lead to an average $log(BF)$ greater than one for a SNP, it significantly impacts the ranking of the SNP; in practice only a few top ranked SNPs receive the (often much more costly) biological experiments. 
Further, the similarity between the two models depends on the sample correlation between the covariates in the two models, $r_{G_1,G_2}$, which in turn depends on the allele frequency of the reference allele.  In the Supplementary Materials, we derive the theoretical correlation, $\rho_{G_1,G_2}$, as a function of allele frequencies in males and females $(p_m,p_f)$, and we show that when $\min\{p_m,p_f\}$ is not close to 1, $G_1$ and $G_2$ are indeed highly correlated.  But when both $p_m$ and $p_f$ are large (e.g.\ $p_m=p_f=0.95$), the correlation can be lower than 0.5.  In that case, the two models will likely lead to different conclusions while the BMA-based approach remains robust.  To illustrate, consider the linear model as before ($n=1000$, $EV=0.01$) but $p_m=p_f=0.95$ and the true simulating model is either $M_1$ or $M_2$.  Table \ref{highAF} shows the average of $\log_{10}BF_{1N}$, $\log_{10}BF_{2N}$ and $\log_{10}BF_{AN}$ obtained from 1000 simulation replicates.  Results here clearly demonstrate the merits of the proposed BMA-based approach. Additional simulations assuming a bigger effect ($EV=0.05$) show that the averages of  $\log_{10}BF_{1N}$, $\log_{10}BF_{2N}$ and $\log_{10}BF_{AN}$ are, respectively,  22.35, 2.29 and 21.65, if the true generating model is $M_1$.
Even if the frequencies are not extreme, say $p_m=p_f=0.3$ as in the application study, the averaged BMA-based quantity is clearly more robust as shown in Table \ref{highAF}, where the BMA-based $\log_{10}BF_{AN}=1.755$, though not as high as $\log_{10}BF_{1N}=1.942$ using the ideal (unknown) true model $M_1$, is clearly a substantial improvement over $\log_{10}BF_{2N}=1.062$ using the incorrect model $M_2$.

When the allele frequency is on the boundary, we have commented that the resulting HPD intervals can be quite wide as seen in the application above e.g.\ $rs12689325$ with MAF of $1.3\%$, the second ranked SNP in Figure \ref{hpd real}; ranked {331} by the minimal p-value approach.   Among the 14220 X-chromosome SNPs analyzed in Section \ref{realD}, 829 SNPs have MAF less than 1\%.  In that case, there is little variation in the genotype variable thus limited information available for inference.  The top ranked SNPs thus were chosen from the remaining 13391 SNPs with MAF greater than 1\%.  In recent years, joint analyses of multiple rare (or common) variants (also known as the gene-based analyses) have received much attention but only for autosome SNPs \citep{andriy14}.  Extension to X-chromosome SNPs remain an open question.  Similarly, additional investigations are needed for X-chromosome SNPs in the areas of family-based association studies \citep{thornton12}, direct interaction studies \citep{cordell09}, as well as indirect interaction studies via scale-test for variance heterogeneity \citep{soave16}.

In Table \ref{coding}, allele $D$ is assumed to be the risk allele with frequency ranging from 0 to 1, i.e.\ not necessary the minor allele.  In practice as in our application,  the minor allele is often assumed to be the risk allele based on the known genetic aetiology for complex traits.  For autosome SNPs,  it is well known that coding $D$ or $d$ as the reference allele leads to identical inference with $\beta_D=-\beta_d$.  However, this is no longer true for analyzing X-chromosome SNPs under the no XCI $M_2$ model assumption; inference is identical under the  XCI $M_1$ model; this was pointed out by \citep{wang14} in their frequentist's approach.  To see the difference empirically in our Bayesian setting, we revisit the CF data as described in Section \ref{realD} and reanalyze SNP $rs3788766$ as a proof of principle. Specifically, we first assume the minor allele $D$ as the reference allele under the XCI and no XCI assumptions, coded respectively as  $M_1$-$D$ and $M_2$-$D$. We then assume $d$ as the reference allele and consider the corresponding $M_1$-$d$ and $M_2$-$d$ models.  Table \ref{switch} shows the Bayes factor $BF_{kN}$ comparing each model with the null model.

It is clear that the XCI $M_1$ model is robust to the choice of the underlying reference allele, i.e.\ $M_1$-$D$ and $M_1$-$d$ lead to identical association inference, but this is not the case for $M_2$-$D$ and $M_2$-$d$ under the assumption of no XCI.  In hindsight this somewhat surprising results can be intuitively explained by the fact that under $M_1$, regardless of the choice of the reference allele, female $dd$ and male $d$ genotypes belong to one group, $DD$ and $D$ belong to another group and $Dd$ itself is a group, i.e.\ $(\{dd, d\}, dD, \{DD, D\})$. Under $M_2$ the three groups are $(\{dd, d\}, \{dD, D\}, DD)$ when $D$ is the reference allele in contrast of  $(dd, \{dD, d\}, \{DD, D\})$ when $d$ is the reference  allele, thus resulting in different association quantities. However, we also note that in practice when allele frequencies are not close to the boundary as in this case (MAF $=0.388$), the empirical difference between $M_2$-$D$ and $M_2$-$d$ is not significant; we provide additional application results in the Supplementary Materials.  Nevertheless, it is worth noting that the choice of reference allele is yet another analytical detail that sets X-chromosome apart from the rest of the genome.

\section*{Acknowledgements}
The authors would like to thank Dr. Lisa J. Strug for providing the cystic fibrosis application data, and Prof. Mike Evans for suggestions that have improved the presentation of the paper.
This research is funded by the Natural Sciences and Engineering Research Council of Canada (NSERC) to RVC and LS, and the Canadian Institutes of Health Research (CIHR) to LS.

\bibliography{mybib}
\bibliographystyle{authordate1}

\begin{table}[h]
\begin{center}
\caption{Genotype coding under the X-chromosome inactivation (XCI, $M_1$) and no X-chromosome inactivation (no XCI, $M_2$) assumptions}
\label{coding}
\begin{tabular}{c|c|c|c|c|c|c} \hline \hline
\multicolumn{2}{c|}{} & \multicolumn{3}{c|}{Female}&\multicolumn{2}{c}{Male}\\ \cline{3-7}
Model & Coding & $dd$ & $dD$ & $DD$ & $d$ & $D$ \\ \hline
$M_1$: XCI  & $G_1$ & 0 & 0.5 & 1 & 0 & 1 \\ \hline
$M_2$: no XCI & $G_2$ & 0 & 1 & 2 & 0 & 1 \\ \hline \hline 
\end{tabular}
\end{center}
\end{table}

\begin{table}[h]
\begin{center}
\caption{Summary of frequentist analysis of four top ranked SNPs, selected from analyzing association evidence between 14220 X-chromosome SNPs and meconium ileus in 3199 Cystic Fibrosis patients.  MAF is the pooled estimate of the frequency of the minor allele (frequencies do not differ between males and females), log odds ratio estimates under the XCI $M_1$ and no XCI $M_2$ assumptions, and their corresponding p-values and the (adjusted) p-value of the minimal p-value or the maximal $Z$ value.}
\label{real1}
\begin{tabular}{c|c|c|c|c|c|c} \hline \hline
& & \multicolumn{2}{c|}{Log Odds Ratios} & \multicolumn{3}{c}{P-values} \\ \cline{3-7}
SNPs & MAF & $M_1$ & $M_2$ & $M_1$ & $M_2$ & $Z_{max}$ \\ \hline
$rs3788766$   & 0.388 & 0.798 & 0.484 & 8.50e-12 & 2.20e-09 & 1.61e-11  \\
$rs5905283$   & 0.487 & 0.586 & 0.326 & 4.79e-08 & 9.64e-06 & 8.88e-08  \\
$rs12839137$ & 0.237 & 0.611 & 0.360 & 7.55e-06 & 5.44e-04 & 1.25e-05  \\
$rs5905284$   & 0.249 & 0.592 & 0.358 & 8.61e-06 & 1.21e-04 & 1.43e-05  \\ \hline \hline 
\end{tabular} 
\end{center}
\end{table}

\begin{table}[h]
\centering
\caption{Summary of Bayesian analysis of 15 top ranked SNPs, selected from analyzing association evidence between 14220 X-chromosome SNPs and meconium ileus in 3199 Cystic Fibrosis patients. SNP are ordered based on their lower bounds of the BMA-average HPD intervals for $\beta$.}
\label{all SNPs}
\begin{tabular}{c|c|c|c|c|c} \hline \hline
  &  & \multicolumn{2}{c|}{HPD intervals} & \multicolumn{2}{c}{Bayes factors}\\ \cline{3-6} 
 SNPs & MAF & Lower & Upper & $BF_{12}$ & $BF_{AN}$ \\ \hline
$rs3788766$ &  0.388 & 0.572 & 1.033 & 2.71e+02 & 1.49e+09 \\ 
  $rs12689325$ & 0.013 & 0.405 & 3.118 & 3.77e -01 & 3.28e+00 \\ 
  $rs5905283$ & 0.487 & 0.379 & 0.784 & 2.02e+02 & 8.71e+04 \\ 
  $rs12845594$ & 0.047 & 0.344 & 1.700 & 7.23e+00 & 8.61e+00 \\ 
  $rs12839137$ & 0.237 & 0.307 & 0.884 & 2.25e+01 & 1.34e+03 \\ 
  $rs5905284$ & 0.249 & 0.302 & 0.830 & 1.83e+01 & 1.08e+03 \\ 
  $rs579854$ & 0.136 & 0.266 & 0.932 & 1.88e+00 & 4.83e+01 \\ 
  $rs5955417$ & 0.030 & 0.260 & 2.130 & 3.28e+00 & 5.30e+00 \\ 
  $rs12720074$ & 0.100 & 0.237 & 0.943 & 5.33e -01 & 5.49e+01 \\ 
  $rs1921965$ & 0.091 & 0.228 & 0.893 & 1.34e+00 & 2.81e+01 \\ 
  $rs6623182$ & 0.036 & 0.217 & 1.216 & 2.89e+00 & 9.08e+00 \\ 
  $rs3027514$ & 0.015 & 0.209 & 1.496 & 8.34e -01 & 1.34e+00 \\ 
  $rs17338514$ & 0.099 & 0.201 & 0.821 & 1.19e+00 & 2.09e+01 \\ 
  $rs11797786$ & 0.068 & 0.191 & 0.947 & 4.68e+00 & 4.47e+00 \\ 
  $rs1921967$ & 0.122 & 0.190 & 0.756 & 7.50e -01 & 2.52e+01 \\ 
   \hline \hline 
\end{tabular}
\end{table}

\begin{table}[h]
\begin{center}
\caption{Summary of the average of $log_{10}$ (Bayes factors) based on 1000 independent simulation replicates. 
Allele frequency of $D$ is 0.95 or 0.3 for both male and female,  sample size is 1000, $EV=0.01$, and  data are simulated from the XCI  $M_1$ model or no XCI $M_2$ model. $BF_{1N}$, $BF_{2N}$ and $BF_{AN}$ are the Bayes factors of  $M_1$ vs. the null $M_N$ model, the no XCI $M_2$ vs. $M_N$, and the BMA-based model vs. $M_N$.}
\label{highAF}
\begin{tabular}{c|c|c|c|c} \hline \hline
$(p_m, p_f)$ & True Model & $\log_{10} BF_{1N}$ & $\log_{10} BF_{2N}$ & $\log_{10} BF_{AN}$ \\ \hline
$(0.95, 0.95)$ & $M_1$ & 2.066 & -1.850 & 1.541  \\
$(0.95, 0.95)$ & $M_2$ & -1.969 & 1.854 & 1.309   \\
$(0.30, 0.30)$    & $M_1$ & 1.942 & 1.062 & 1.755  \\
$(0.30, 0.30)$    & $M_2$ & 1.073 & 1.983 & 1.796  \\ \hline \hline
\end{tabular} 
\end{center}
\end{table}

\begin{table}[h]
\begin{center}
\caption{Bayes factors of comparing models assuming different risk reference allele ($D$ or $d$) and genotype codings (XCI $M_1$ or no XCI $M_2$) of SNP $rs3788766$ with the null model, based on the Cystic Fibrosis application data.}
\label{switch}
\begin{tabular}{c|c|c|c|c|c|c} \hline \hline
& \multicolumn{3}{c|}{Female}&\multicolumn{2}{c|}{Male} & \\ \cline{2-4} \cline{5-6} 
Model & dd & dD & DD & d & D & $BF_{kN}$ \\ \hline
$M_1$-$D$  & 0 & 0.5 & 1 & 0 & 1 & 2.97e+09\\
$M_1$-$d$ & 1 & 0.5 & 0 & 1 & 0 & 2.97e+09\\
$M_2$-$D$ & 0 & 1 & 2 & 0 & 1 & 8.21e+06\\ 
$M_2$-$d$ & 2 & 1 & 0 & 1 & 0 & 1.74e+06\\ \hline \hline
\end{tabular}
\end{center}
\end{table}

\end{document}